\documentclass[12pt,usenames,dvipsnames]{article}

\usepackage{ragged2e}
\usepackage{sectsty}

\usepackage{amsmath,amsthm,amssymb,amscd,mathtools,bm,upgreek}
\usepackage{graphicx}
\usepackage{epsfig}
\usepackage{hyperref}
\hypersetup{colorlinks,linkcolor=blue,citecolor=blue,urlcolor = blue}
\usepackage{rotating}
\usepackage{float}
\usepackage{enumitem}
\usepackage{multirow}
\usepackage{booktabs}
\usepackage{eurosym}
\usepackage{caption}
\usepackage{pdfcolfoot}
\usepackage{longtable}
\usepackage[title]{appendix}
\RequirePackage{mathtools,algpseudocode,etoolbox}
\usepackage[linesnumbered,ruled,vlined]{algorithm2e}

\usepackage{listings}
\usepackage{color} 
\usepackage[style=apa,backend=biber]{biblatex}
\DeclareSourcemap{
  \maps[datatype=bibtex]{
    \map{
      \step[fieldset=issn, null]
      \step[fieldset=isbn, null]
      \step[fieldset=doi, null]
      \step[fieldset=url, null]
      \step[fieldset=note, null]
    }
  }
}
\usepackage{subcaption}
\usepackage{mwe}

\usepackage{siunitx}

\usepackage{xcolor,soul}

\usepackage{authblk}

\def\T{{\mathcal{T}}}

\def\vy{{\boldsymbol{y}}}
\def\vx{{\boldsymbol{x}}}

\def\vu{{\boldsymbol{u}}}
\def\vv{{\boldsymbol{v}}}

\def\vbeta{{\boldsymbol \beta}}
\def\vgamma{{\boldsymbol \gamma}}

\def\reals{\mathbb{R}}

\def\comp{\raise 1pt \hbox{$\scriptstyle\circ$}}

\def\upto{{\raise 1pt \hbox{$\scriptstyle \,\nearrow\,$}}}
\def\downto{{\raise 1pt \hbox{$\scriptstyle \,\searrow\,$}}}

\title{\Large\textbf{Axiomatic modeling of fixed proportion technologies}
}

\author{
\begin{tabular}{cc}
Xun Zhou\thanks{%
\textit{Corresponding author.} 
Surrey Business School, University of Surrey, UK. Email: \href{mailto:x.zhou@surrey.ac.uk}{x.zhou@surrey.ac.uk}.}
\hspace{.5em}
Timo Kuosmanen\thanks{%
Turku School of Economics, University of Turku, Finland. Email: \href{mailto:timo.kuosmanen@utu.fi}{timo.kuosmanen@utu.fi}.} 
\end{tabular}
}

\date{November 2024}

\begin{document}
\captionsetup[figure]{labelfont={bf},labelformat={default},labelsep=period,name={Figure},font=small}
\captionsetup[table]{labelfont={bf},labelformat={default},labelsep=period,name={Table},font=small}

\maketitle

\vfill
\begin{center}
Declarations of interest: none
\end{center}
\vfill

\begin{abstract}
\noindent 
Understanding input substitution and output transformation possibilities is critical for efficient resource allocation and firm strategy. There are important examples of fixed proportion technologies where certain inputs are non-substitutable and/or certain outputs are non-transformable. However, there is widespread confusion about the appropriate modeling of fixed proportion technologies in data envelopment analysis. We point out and rectify several misconceptions in the existing literature, and show how fixed proportion technologies can be correctly incorporated into the axiomatic framework. A Monte Carlo study is performed to demonstrate the proposed solution.
\\[5mm]
\\[2mm]
\end{abstract}
\vfill

\thispagestyle{empty}
\newpage
\setcounter{page}{1}
\setcounter{footnote}{0}
\pagenumbering{arabic}
\baselineskip 20pt
\setlength\bibitemsep{1.15\itemsep}

\section{Introduction}\label{sec:intro}
The Leontief production function, which forms the basis for the input-output analysis \parencite{leontief_structure_1941}, is a classic parametric specification of the production function in economics (e.g., \cite{diewert_application_1971}). The Leontief production function is also known as the fixed proportions production function because it assumes the inputs must be used in specific technologically predetermined proportions; in other words, there is no substitutability between inputs. There are several important examples of production technologies where certain inputs are non-substitutable and/or certain outputs are non-transformable (e.g., \cite{Barnum2011,Boyabatl2015,guner_multi-period_2021}).

The axiomatic theory of production that builds upon such works as \textcite{Koopmans1951}, \textcite{Farrell1957}, \textcite{Shephard1970}, and \textcite{Afriat1972} is nowadays commonly referred to as data envelopment analysis (DEA) \parencite{Charnes1978}. DEA does not generally assume anything about substitutability (transformability) among inputs (outputs). However, the empirical DEA frontiers are piece-wise linear surfaces that tend to exhibit at least limited substitution (transformation) possibilities. 

To improve the DEA estimator in the case of fixed proportion technologies, \textcite{Barnum2011} and \textcite{Barnum2017} propose extended DEA formulations that are claimed to impose non-substitutability. According to \textcite{Barnum2017}, failure to impose non-substitutability is ``a ubiquitous methodological hazard in DEA modelling". \textcite{barnum_bias_2021} further recommend that ``DEA studies should test for substitution and transformation in their data, to avoid biased DEA scores and spurious second-stage regression results".

The first contribution of this paper is to demonstrate that, in fact, the proposed solution by \textcite{Barnum2011} imposes perfect substitution instead of the intended non-substitutability. We believe it is important to clarify this rather fundamental misconception in order to strengthen the axiomatic foundation of DEA and avoid misleading results in applications.

The second contribution of this paper is to demonstrate how the fixed proportion technologies can be correctly incorporated into the axiomatic framework. We present a simple branch-and-cut algorithm that is straightforward to implement in any software or computational package.

Our third contribution is to present Monte Carlo evidence in support of our arguments. Our Monte Carlo simulations illustrate that standard DEA that ignores non-substitutability remains consistent even when the inputs are not substitutable, however, ignoring non-substitutability increases finite sample bias. The simulations demonstrate that our proposed solution helps to decrease finite sample bias when the assumption of non-substitutability holds, however, \citeauthor{Barnum2011}'s (\citeyear{Barnum2011}) solution only makes things worse.

The rest of the paper is organized as follows. Section \ref{sec:setup} introduces the classic axioms and introduces formally the new axioms of non-substitutability of inputs and non-transformability of outputs. Section \ref{sec:corr} demonstrates how the proposed axioms can be correctly implemented in DEA. The Monte Carlo simulations and their results are presented and discussed in Section \ref{sec:mc-sim}. Finally, Section \ref{sec:concl} presents our concluding remarks. 

\section{Axiomatic approach}\label{sec:setup}
Consider $N$ decision-making units (DMUs). Each DMU operates a joint production technology that transforms $M$ inputs $\vx_n = (x_{1n}, \ldots, x_{Mn}) \in \reals_+^M$ to $S$ outputs $\vy_n = (y_{1n}, \ldots, y_{Sn}) \in \reals_+^S$, $n = 1, \ldots, N$. The most obvious representation of the production technology is the production possibility set $\T = \bigl\{(\vx,\vy)\,|\,\vx\;{\rm{can}}\,{\rm{produce}}\;\vy\bigl\}$.

The three classical axioms of the production technology considered in this paper include the following (see, e.g., \cite{Koopmans1951,Afriat1972,Fare1995}):
\begin{itemize}
    \item[] A1. Free disposability of inputs and outputs: If $(\vx,\vy) \in \T$ and $\tilde{\vx} \geq \vx, \tilde{\vy} \leq \vy$, then $(\tilde{\vx},\tilde{\vy}) \in \T$.
    \item[] A2. Convexity: If $(\vx,\vy) \in \T$, then $\left( \sum\limits_{n=1}^N \lambda_n\vx_n, \sum\limits_{n=1}^N \lambda_n\vy_n \right) \in \T$ for any scalar $\lambda \ge 0$ such that $\sum\limits_{n=1}^N \lambda_n =1$.
    \item[] A3. Constant returns to scale: If $(\vx,\vy) \in \T$, then $(a\vx,a\vy) \in \T$ for any scalar $a > 0$.
\end{itemize}

A subset of non-dominated $(\vx,\vy)$ in $\T$ relevant to this paper is the weakly efficient subset, which can be defined as follows (see, e.g., \cite{fare_structure_1983,kuosmanen_dea_2001,mehdiloo_strong_2021}):
\begin{equation*}
    \,\text{WEff}\,(\T) = \bigl\{ (\vx,\vy) \in \T\,|\,\tilde{\vx} < \vx, \tilde{\vy} > \vy \Rightarrow (\tilde{\vx},\tilde{\vy}) \notin \T \bigl\}.
\end{equation*}
Note that if $\T$ is convex (i.e., A2 holds), then for all $(\vx,\vy)$ in $\text{WEff}\,(\T)$, there exists a supporting hyperplane $H(\vx,\vy) =
\alpha + \vbeta'\vx + \vgamma'\vy$ containing $(\vx,\vy)$ such that $\T$ is entirely contained in one of the two closed half-spaces bounded by $H(\vx,\vy)$. This result is known as the supporting hyperplane theorem. However, the supporting hyperplane $H(\vx,\vy)$ is not necessarily unique.

In a special case where a pair of inputs are non-substitutable, we may introduce an additional assumption in terms of $\T$ after the maintained axioms:
\begin{itemize}
    \item[] A4. Non-substitutability of inputs $m$ and $\tilde{m}$: For any $(\vx,\vy)$ in $\text{WEff}\,(\T)$, there exists a supporting hyperplane $H(\vx,\vy)$ such that $\beta_m=0$ or $\beta_{\tilde{m}}=0$.
\end{itemize}
Analogously, if a pair of outputs are non-transformable, we may introduce another additional assumption:
\begin{itemize}
    \item[] A5. Non-transformability of outputs $s$ and $\tilde{s}$: For any $(\vx,\vy)$ in $\text{WEff}\,(\T)$, there exists a supporting hyperplane $H(\vx,\vy)$ such that $\gamma_s=0$ or $\gamma_{\tilde{s}}=0$.
\end{itemize}
Imposing either A4 or A5 as an additional assumption results in a fixed proportion technology.

To illustrate the new axioms A4 and A5, consider the classic \textcite{leontief_structure_1941} production function in a two-input case: 
\begin{equation*}
    y=\min ({x_1}/{a},{x_2}/{b}),
\end{equation*}
where $a$ and $b$ are constants determined by the technology \parencite{diewert_application_1971}. Note that this production function is equivalent to the fixed proportion technology
\begin{equation*}
    \T^L = \bigl\{(\vx,y)\,|\,y \leq \min({x_1}/{a},{x_2}/{b}) \bigl\}.
\end{equation*}

\begin{figure}[H]
    \centering
    \includegraphics[width=0.65\linewidth]{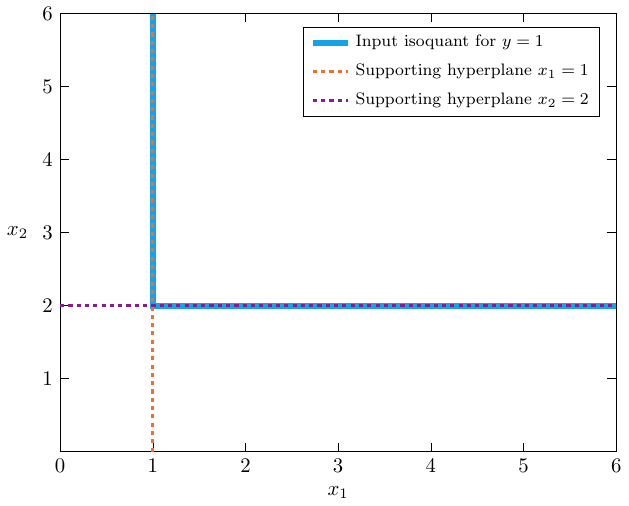}
    \caption{Input isoquant and supporting hyperplanes for a Leontief production function with $y=1$, $a=1$, and $b=2$.}
    \label{fig:leontief}
\end{figure}

Figure \ref{fig:leontief} plots the L-shaped input isoquant of this technology, assuming the output level $y=1$ and parameter values $a=1$ and $b=2$. Note that the cost-minimizing input vector is $(x_1, x_2)=(1,2)$ at any non-negative input prices. Note further that if one of the inputs is fixed at the cost-minimizing level, then any increase in the other input beyond the cost-minimizing level does not increase the output. 

Figure \ref{fig:leontief} also plots two supporting hyperplanes along the horizontal and vertical segments, respectively. For point $(x_1, x_2)=(1,2)$, there is no unique supporting hyperplane: the set of supporting hyperplanes is bounded by the two hyperplanes plotted in Figure \ref{fig:leontief}. For all other points on the isoquant, there is only one supporting hyperplane, which is one of the two hyperplanes plotted in Figure \ref{fig:leontief}. 

The horizontal supporting hyperplane is
\begin{equation*}
    H(x_1,x_2,y)=0x_1+\frac{1}{2}x_2-y=0,
\end{equation*}
which implies $\beta_1 = 0$ and $\beta_2 = \frac{1}{2}$. Analogously, the vertical supporting hyperplane is
\begin{equation*}
    H(x_1,x_2,y)=x_1+0x_2-y=0,
\end{equation*}
implying $\beta_1 = 1$ and $\beta_2 = 0$. Hence, we have shown that for every $(x_1, x_2)$ on the isoquant, including point $(x_1, x_2)=(1,2)$, there exists a supporting hyperplane $H(x_1,x_2,y)$ such that $\beta_1 = 0$ or $\beta_2 = 0$. This argument does not depend on the arbitrary choice of output $y$ or parameter values $a,b$. Therefore, the fixed proportion technology $\T^L$ satisfies axiom A4.

It is worth noting that A4 (A5) focuses on the non-substitutability (non-transformability) between a pair of inputs (outputs). It is possible to generalize these axioms to an arbitrary subset of non-substitutable inputs and/or non-transformable outputs, or even multiple subsets of non-substitutable inputs and/or non-transformable outputs. We leave such generalizations as an interesting topic for future research.



\section{Correct implementation}\label{sec:corr}
\textcite{Afriat1972} was the first to introduce and prove the minimum extrapolation production functions in the single output case for the following combinations of axioms: A1, \{A1, A2\}, and \{A1, A2, A3\}. Extensions to the multiple output setting were presented for A1 in \textcite{tulkens_fdh_1993} known as the free disposal hull (FDH) technology, for \{A1, A2\} in \textcite{banker1984} known as the variable returns to scale (VRS) or BCC technology, and for \{A1, A2, A3\} in \textcite{Charnes1978} known as the constant returns to scale (CRS) or CCR technology. The multi-output DEA technologies can be stated in terms of $\T$ as follows:
\begin{itemize}
    \item[] FDH: $\,\T_{\text{A1}}^{\,\text{DEA}} = \Bigl\{ (\vx,\vy)\,\Big|\,\sum\limits_{n=1}^N \lambda_n\vx_n \le \vx;\sum\limits_{n=1}^N \lambda_n\vy_n \ge \vy;\sum\limits_{n=1}^N \lambda_n =1;\lambda_n \in \{0,1\},\forall n \Bigl\}$.
    \item[] BCC: $\,\T_{\text{A1,A2}}^{\,\text{DEA}} = \Bigl\{ (\vx,\vy)\,\Big|\,\sum\limits_{n=1}^N \lambda_n\vx_n \le \vx;\sum\limits_{n=1}^N \lambda_n\vy_n \ge \vy;\sum\limits_{n=1}^N \lambda_n =1;\lambda_n \ge 0,\forall n \Bigl\}$.
    \item[] CCR: $\,\T_{\text{A1,A2,A3}}^{\,\text{DEA}} = \Bigl\{ (\vx,\vy)\,\Big|\,\sum\limits_{n=1}^N \lambda_n\vx_n \le \vx;\sum\limits_{n=1}^N \lambda_n\vy_n \ge \vy;\lambda_n \ge 0,\forall n \Bigl\}$.
\end{itemize}
To maintain clarity and brevity, we will focus on the input-oriented CCR technology in the subsequent discussion. Nonetheless, our findings can readily be extended to the BCC and FDH technologies as well as other orientations. 

Having introduced the classic envelopment formulations, it becomes evident that it is more straightforward to implement A4 and A5 using the multiplier formulation. This formulation allows for the explicit incorporation of non-substitutability of inputs and non-transformability of outputs through additional constraints on the input and output weights \parencite{Allen1997}. The DEA technology frontier can be estimated by the CCR model in its multiplier form as follows:
\begin{equation}\label{eq:CCR}
\max_{\vu_n, \vv_n} \,\ \theta_{n}=\sum_{s=1}^S u_{sn} y_{sn}
\end{equation}
\begin{equation*}
\mbox{\textit{s.t.}}\quad
\sum_{m=1}^M v_{mn} x_{mn}=1,    
\end{equation*}
\begin{equation*}
\sum_{s=1}^S u_{sn} y_{sj}-\sum_{m=1}^M v_{mn} x_{nj} \leq 0, \quad j = 1, \ldots, N,
\end{equation*}
\begin{equation*}
u_{sn} \ge 0,v_{mn} \ge 0, \quad s = 1, \ldots, S, \; m = 1, \ldots, M.
\end{equation*}
where $\theta_{n}$ is a scalar denoting the technical efficiency, $\vv_n = (v_{1n}, \ldots, v_{Mn}) \in \reals_+^M$ is a vector of the input weights, and $\vu_n = (u_{1n}, \ldots, u_{Sn}) \in \reals_+^S$ is a vector of the output weights for DMU $n$, $n = 1, \ldots, N$.

If one has prior information and is willing to make additional assumptions that a pair of inputs are non-substitutable (A4) and/or a pair of outputs are non-transformable (A5), we can incorporate those as extra constraints into the CCR multiplier form:
\begin{equation}\label{eq:FP-CCR}
\max_{\vu_n, \vv_n} \,\ \theta_{n}=\sum_{s=1}^S u_{sn} y_{sn}
\end{equation}
\begin{equation*}
\mbox{\textit{s.t.}}\quad
\sum_{m=1}^M v_{mn} x_{mn}=1,    
\end{equation*}
\begin{equation*}
\sum_{s=1}^S u_{sn} y_{sj}-\sum_{m=1}^M v_{mn} x_{nj} \leq 0, \quad j = 1, \ldots, N,
\end{equation*}
\begin{equation*}
u_{sn} \ge 0,v_{mn} \ge 0, \quad s = 1, \ldots, S, \; m = 1, \ldots, M,
\end{equation*}
\begin{equation*}
\text{either}\; v_{mn}=0 \;\text{or}\; v_{\tilde{m}n}=0, \exists m,\tilde{m} \in \{1,\ldots, M\},\, m \neq \tilde{m},
\end{equation*}
\begin{equation*}
\text{either}\; u_{sn}=0 \;\text{or}\; u_{\tilde{s}n}=0, \exists s,\tilde{s} \in \{1,\ldots, S\},\, s \neq \tilde{s}.
\end{equation*}
The formulation of problem \eqref{eq:FP-CCR} is the same as problem \eqref{eq:CCR} except for the final two sets of constraints. The second last set imposes assumption A4, enforcing the marginal rate of substitution (MRS) between the pair of non-substitutable inputs to be either zero or infinite. The last set of constraints imposes A5, enforcing the marginal rate of transformation (MRT) between the pair of non-transformable outputs to be either zero or infinite.

Problem \eqref{eq:FP-CCR} is a disjunctive programming problem, which is computationally demanding due to the need for specialized optimization solvers. Fortunately, there exists a simple branch-and-cut algorithm that breaks down problem \eqref{eq:FP-CCR} into a stepwise procedure. This method, which we will refer to as FP-constrained DEA, offers a more accessible approach. Without the loss of generality, consider a basic scenario with two inputs and a single output, where the two inputs are non-substitutable. Based on this scenario, the stepwise procedure of FP-constrained DEA is demonstrated in Algorithm \ref{alg:FP-DEA}.

\vspace{.45em}
\begin{algorithm}[H]\label{alg:FP-DEA} 
  \KwData{$\{\vx_n, y_n\}_{n=1}^N \in \reals_+^2 \times \reals_+$}
  $\theta_{n} = 0, n = 1, \ldots, N$\;
  \While{$\theta_{n} = 0$}{
  	Solve problem \eqref{eq:CCR} excluding $x_{1n}$ while keeping all other           variables to calculate $\theta_{n}^1$;
        
        Solve problem \eqref{eq:CCR} excluding $x_{2n}$ while keeping all other variables to calculate $\theta_{n}^2$;
    \eIf{$\theta_{n}^1\ge \theta_{n}^2$}{
    $\theta_{n} = \theta_{n}^1$\;
    }{
    $\theta_{n} = \theta_{n}^2$\;
    }
 }
 \KwResult{$\theta_{n}$}
\caption{Calculating technical efficiency $\theta_{n}$ using FP-constrained DEA.}
\end{algorithm}
\vspace{1em}
\section{Monte Carlo simulations}\label{sec:mc-sim}
Consider a fixed proportion technology with $M$ inputs and a single output, where the true production function is $F(\vx_n) = \min\{x_{1n}, \ldots, x_{Mn}\}, n = 1, \ldots, N$. The inputs are randomly generated from the uniform distribution in the range $[0, 100]$. The observed output $y_n = F(\vx_n)\times \exp(-\mu_n)$ is perturbed by a half-normal inefficiency term $\mu_n\geq 0$ drawn independently from $\mathcal{N}^+(0,\sigma_\mu^2)$. No stochastic noise is considered as standard DEA does not handle it. The fixed proportion technology is assumed to exhibit CRS, and the technology frontier is characterized by the Shephard input distance function. Once the frontier is estimated, it is possible to measure efficiency using various types of distance metrics or slack-based measures, but this is another question that is not dependent on frontier estimation. In this study, we do not consider non-radial slacks but simply use the radial Farrell measure of technical efficiency.

We examine 48 scenarios encompassing input dimensions $M \in \{2, 3\}$, no inefficiency and 3 levels of inefficiency with standard deviations $\sigma_\mu \in \{1, 2, 3\}$, along with 6 different sample sizes $N \in \{30, 50, 100, 300, 500, 1000\}$. Each scenario is replicated 1000 times using the DEA toolbox developed by \textcite{alvarez_data_2020} on MATLAB R2021a. The computation was undertaken on the Viking Cluster, a high performance computing facility provided by the University of York. 

Since we know the true distance to the frontier of the fixed proportion technology, we can compare the finite sample performance of the original CCR DEA and our proposed FP-constrained DEA relative to the true frontier. In addition, we include \citeauthor{Barnum2011}'s (\citeyear{Barnum2011}) proposed modeling of fixed proportion technologies in this comparison. \textcite{Barnum2011} and \textcite{Barnum2017} impose identical input (output) weights between any pair of non-substitutable inputs (non-transformable outputs) in the CCR and BBC DEA formulations, respectively. Technically, perfect substitutability (transformability) between inputs (outputs) is a direct consequence of this modeling approach. In other words, their results would be exactly the same if perfect substitutability (transformability) was assumed. 

A standard performance measure is the mean squared error (MSE) between the true and estimated distance to the frontier (technical efficiency scores). The smaller the MSE, the better the finite sample performance. Table \ref{tab:mc-mse1} presents the MSE statistics for different methods across diverse scenarios. 

\begin{table}[H]
   \caption{MSE statistics for CCR DEA (CCR), FP-constrained DEA (FP), and \citeauthor{Barnum2011}'s (\citeyear{Barnum2011}) proposal (BG).}
\renewcommand\arraystretch{1}
\footnotesize{%
\centerline{
\begin{tabular}{ccccccccccccc}
\toprule
\multirow{2}{*}{$M$} & \multirow{2}{*}{$N$} & \multicolumn{3}{c}{$\sigma_\mu=1$} & \multicolumn{3}{c}{$\sigma_\mu=2$} & \multicolumn{3}{c}{$\sigma_\mu=3$} & \multicolumn{2}{c}{no inefficiency} \\ \cmidrule{3-13} 
  &        & CCR      & FP  &  BG     & CCR      & FP  &  BG   & CCR      & FP &  BG    & CCR \& FP  &  BG     \\ \midrule
2 & 30     & 0.0076     & 0.0035  & 0.0619  & 0.0120     & 0.0068  & 0.0344  & 0.0146     & 0.0093  & 0.0248  & 0    & 0.2230                           \\
  & 50     & 0.0038     & 0.0014  & 0.0631  & 0.0061     & 0.0028  & 0.0340  & 0.0076     & 0.0041  & 0.0234  & 0    & 0.2231                         \\
  & 100    & 0.0013     & 0.0004  & 0.0668  & 0.0021     & 0.0008  & 0.0358  & 0.0026     & 0.0011  & 0.0239  & 0    & 0.2251                            \\
  & 300    & 0.0003     & 0.0000  & 0.0701  & 0.0004     & 0.0001  & 0.0382  & 0.0006     & 0.0002  & 0.0256   & 0    & 0.2265                              \\
  & 500    & 0.0001     & 0.0000  & 0.0717  & 0.0002     & 0.0000  & 0.0392  & 0.0002     & 0.0001  & 0.0263   & 0    & 0.2268                        \\
  & 1000   & 0.0000     & 0.0000  & 0.0732  & 0.0001     & 0.0000  & 0.0403  & 0.0001     & 0.0000  & 0.0273   & 0    & 0.2270                             \\
3 & 30     & 0.0189     & 0.0071  & 0.0890  & 0.0261     & 0.0136  & 0.0485  & 0.0291     & 0.0177  & 0.0340   & 0    & 0.3311                            \\
  & 50     & 0.0097     & 0.0029  & 0.0916   & 0.0134     & 0.0056  & 0.0485  & 0.0153     & 0.0076  & 0.0325   & 0    & 0.3360                            \\
  & 100    & 0.0042     & 0.0008  & 0.0985   & 0.0058     & 0.0017  & 0.0525  & 0.0068     & 0.0024  & 0.0349   & 0    & 0.3439                          \\
  & 300    & 0.0010     & 0.0001  & 0.1056  & 0.0014     & 0.0002  & 0.0569  & 0.0017     & 0.0003  & 0.0379   & 0    & 0.3519                           \\
  & 500    & 0.0005     & 0.0000  & 0.1081   & 0.0007     & 0.0001  & 0.0586  & 0.0009     & 0.0001  & 0.0392   & 0    & 0.3549                           \\
  & 1000   & 0.0002     & 0.0000  & 0.1109  & 0.0003     & 0.0000  & 0.0607   & 0.0004     & 0.0000  & 0.0409   & 0   & 0.3568                           \\ 
\bottomrule
\end{tabular}
}}
\label{tab:mc-mse1}
\end{table}

In the case of no inefficiency, both CCR DEA and FP-constrained DEA achieve a perfect fit for all input dimensions and sample sizes, whereas \citeauthor{Barnum2011}'s (\citeyear{Barnum2011}) proposal yields estimates noticeably deviating from the true technical efficiency scores. In the presence of inefficiencies, CCR DEA demonstrates strong performance in estimating fixed proportion technologies, as evidenced by low MSE values. Impressively, FP-constrained DEA outperforms even further, achieving MSE values that are closer to zero. In sharp contrast, \citeauthor{Barnum2011}'s (\citeyear{Barnum2011}) proposal exhibits notably diminished performance. These findings are further confirmed by the correlation coefficients between the true distance to the frontier and the estimates, as reported in Table \ref{tab:mc-cor1}. More detailed findings in the presence of inefficiencies are summarized as follows:
\begin{itemize}
    \item The finite sample performance of both CCR DEA and FP-constrained DEA becomes slightly worse as the standard deviation of inefficiency increases or an additional input is included.
    \item As the sample size increases, both CCR DEA and FP-constrained DEA exhibit reduced MSE values and increased correlation coefficients, thus signifying enhanced finite sample performance.
    \item FP-constrained DEA, in comparison to CCR DEA, achieves consistently better finite sample performance, particularly in small samples. 
\end{itemize}

Overall, the Monte Carlo study has provided numerical evidence 1) that the original CCR DEA modeling of fixed proportion technologies suffers from finite sample bias but approaches the true frontier as the sample size increases and 2) that our proposed FP-constrained DEA approach offers more accurate modeling of fixed proportion technologies and alleviates finite sample bias, particularly in small samples.

\begin{table}[H]
   \caption{Correlation coefficients for CCR DEA (CCR), FP-constrained DEA (FP), and \citeauthor{Barnum2011}'s (\citeyear{Barnum2011}) proposal (BG).}
\renewcommand\arraystretch{1}
\footnotesize{%
\centerline{
\begin{tabular}{ccccccccccc}
\toprule
\multirow{2}{*}{$m$} & \multirow{2}{*}{$N$} & \multicolumn{3}{c}{$\sigma_u=1$} & \multicolumn{3}{c}{$\sigma_u=2$} & \multicolumn{3}{c}{$\sigma_u=3$} \\ \cmidrule{3-11} 
 &                                                                        & CCR      & FP  & BG    & CCR      & FP  & BG   & CCR      & FP & BG     \\ \midrule
2                  & 30                                                                     & 0.9862        & 0.9941     &  0.6836    & 0.9859        & 0.9920    & 0.8092    & 0.9839        & 0.9896     & 0.8501   \\
                   & 50                                                                     & 0.9917        & 0.9976  & 0.6865      & 0.9914        & 0.9964   & 0.8121     & 0.9901        & 0.9949    & 0.8521    \\
                   & 100                                                                    & 0.9967        & 0.9993  & 0.6887      & 0.9965        & 0.9989   & 0.8131     & 0.9958        & 0.9984    & 0.8524     \\
                   & 300                                                                    & 0.9990        & 0.9999  & 0.6878      & 0.9989        & 0.9998   & 0.8123     & 0.9987        & 0.9997   & 0.8514     \\
                   & 500                                                                    & 0.9995        & 1.0000  & 0.6883      & 0.9994        & 0.9999   & 0.8126     & 0.9993        & 0.9999   & 0.8516     \\
                   & 1000                                                                   & 0.9998        & 1.0000  & 0.6875       & 0.9998        & 1.0000   & 0.8120     & 0.9997        & 1.0000   & 0.8511     \\
3                  & 30                                                                     & 0.9608        & 0.9844   & 0.5944     & 0.9645        & 0.9795   & 0.7422     & 0.9628        & 0.9750    & 0.7949    \\
                   & 50                                                                     & 0.9770        & 0.9935   & 0.6057     & 0.9797        & 0.9910   & 0.7479     & 0.9782        & 0.9882    & 0.7977    \\
                   & 100                                                                    & 0.9882        & 0.9979   & 0.6009     & 0.9892        & 0.9968   & 0.7421     & 0.9882        & 0.9955  & 0.7909       \\
                   & 300                                                                    & 0.9961        & 0.9997   & 0.6027     & 0.9963        & 0.9995   & 0.7433     & 0.9959        & 0.9993 & 0.7916        \\
                   & 500                                                                    & 0.9977        & 0.9999   & 0.6023     & 0.9978        & 0.9998   & 0.7429     & 0.9975        & 0.9997 & 0.7912        \\
                   & 1000                                                                   & 0.9989        & 1.0000   & 0.6030      & 0.9989        & 1.0000   & 0.7433     & 0.9988        & 0.9999  & 0.7914       \\
\bottomrule
\end{tabular}
}}
\vspace{0.5em}
\footnotesize{Note: Correlation coefficients in the case of no inefficiency are undefined as the true technical efficiency scores consistently equal 1.}
\label{tab:mc-cor1}
\end{table}

\begin{figure}[H]
	\centering
	\begin{subfigure}[b]{0.49\textwidth}
		\centering
		\includegraphics[width=1\textwidth]{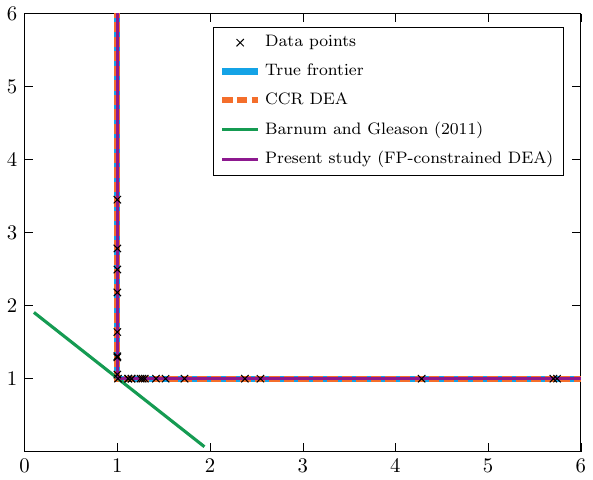} 
		\caption[]%
		{{\small no inefficiency}}    
		\label{fig30:a}
	\end{subfigure}
	\begin{subfigure}[b]{0.49\textwidth}  
		\centering 
		\includegraphics[width=1\textwidth]{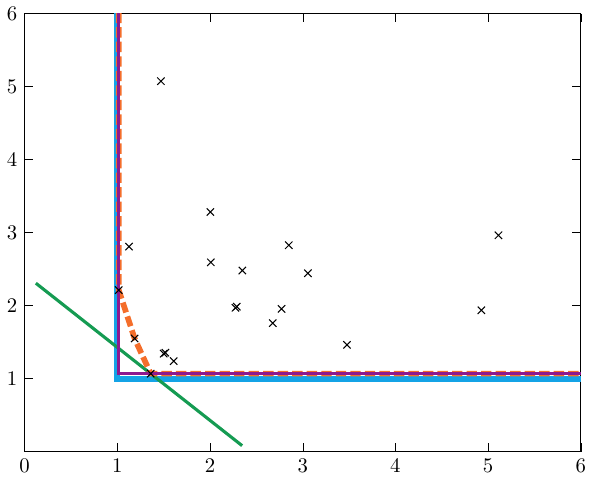} 
		\caption[]%
		{{\small $\sigma_\mu=1$}}    
		\label{fig30:b}
	\end{subfigure}
	\begin{subfigure}[b]{0.49\textwidth}   
		\centering 
		\includegraphics[width=1\textwidth]{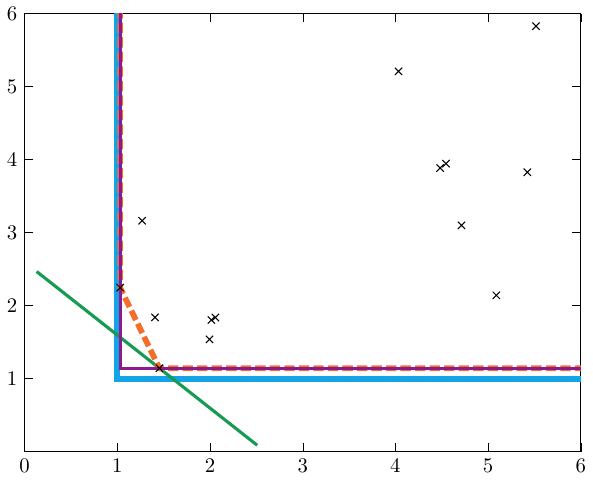} 
		\caption[]%
		{{\small $\sigma_\mu=2$}}    
		\label{fig30:c}
	\end{subfigure}
	\begin{subfigure}[b]{0.49\textwidth}   
		\centering 
		\includegraphics[width=1\textwidth]{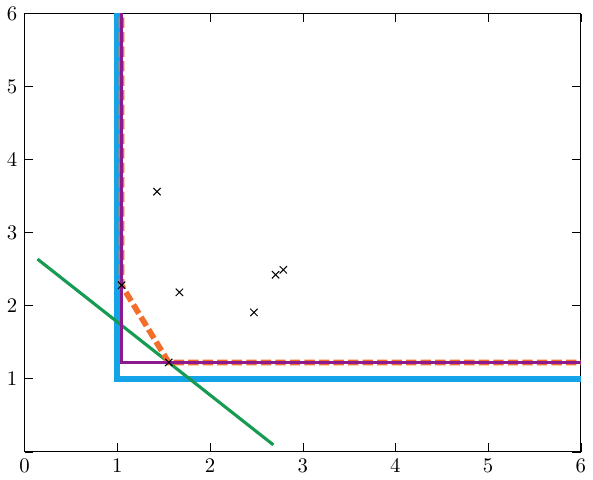} 
		\caption[]%
		{{\small $\sigma_\mu=3$}}    
		\label{fig30:d}
	\end{subfigure}
	\caption[]%
	{Illustration of the input isoquants estimated by CCR DEA, FP-constrained DEA, and \citeauthor{Barnum2011}'s (\citeyear{Barnum2011}) proposal with $N=30$. X-axis: $x_1/y$, Y-axis: $x_2/y$.}
	\label{fig:sample30}
\end{figure}

\begin{figure}[H]
	\centering
	\begin{subfigure}[b]{0.49\textwidth}
		\centering
		\includegraphics[width=1\textwidth]{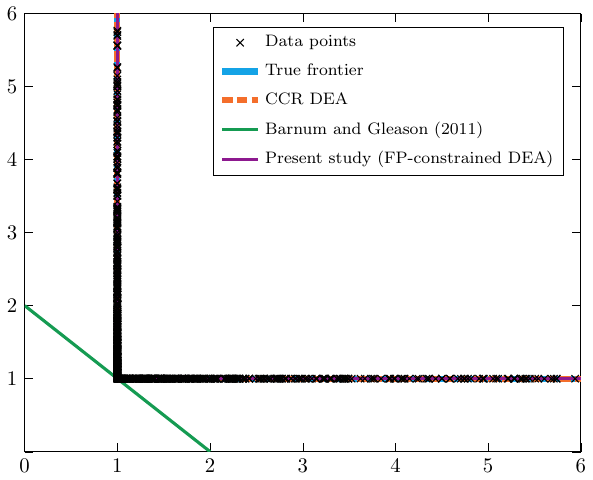} 
		\caption[Network2]%
		{{\small no inefficiency}}    
		\label{fig1000:a}
	\end{subfigure}
	\begin{subfigure}[b]{0.49\textwidth}  
		\centering 
		\includegraphics[width=1\textwidth]{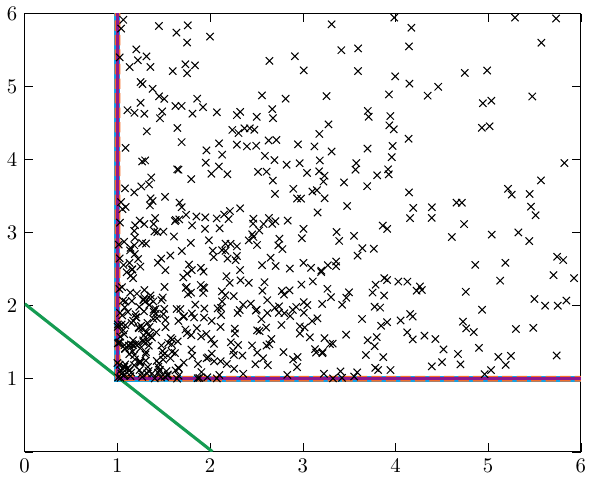} 
		\caption[]%
		{{\small $\sigma_\mu=1$}}    
		\label{fig1000:b}
	\end{subfigure}
	\begin{subfigure}[b]{0.49\textwidth}   
		\centering 
		\includegraphics[width=1\textwidth]{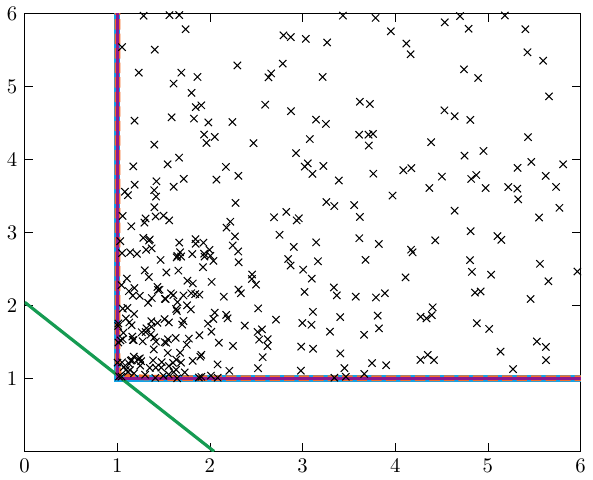} 
		\caption[]%
		{{\small $\sigma_\mu=2$}}    
		\label{fig1000:c}
	\end{subfigure}
	\begin{subfigure}[b]{0.49\textwidth}   
		\centering 
		\includegraphics[width=1\textwidth]{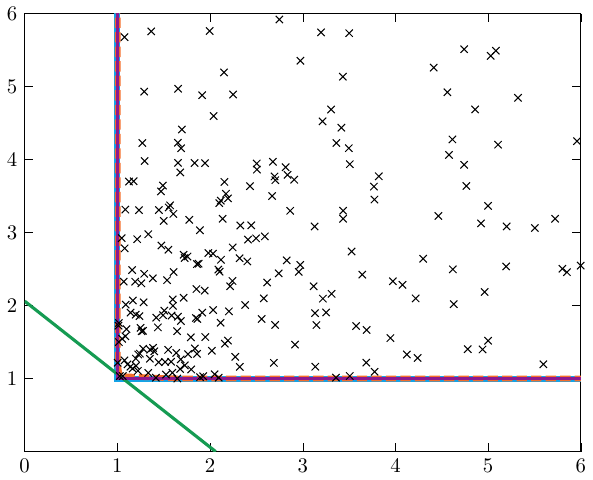} 
		\caption[]%
		{{\small $\sigma_\mu=3$}}    
		\label{fig1000:d}
	\end{subfigure}
	\caption[]%
	{Illustration of the input isoquants estimated by CCR DEA, FP-constrained DEA, and \citeauthor{Barnum2011}'s (\citeyear{Barnum2011}) proposal with $N=1000$. X-axis: $x_1/y$, Y-axis: $x_2/y$.}
	\label{fig:sample1000}
\end{figure}

To gain more insight, we proceed to plot the true input isoquants and those estimated by CCR DEA and FP-constrained DEA, as well as \citeauthor{Barnum2011}'s (\citeyear{Barnum2011}) proposal, based on random samples of the two extreme sample sizes in this study (30 and 1000) within the scenarios of two inputs. Figures \ref{fig:sample30} and \ref{fig:sample1000} illustrate the four sets of input isoquants with the smallest (30) and largest (1000) sample sizes, respectively.

Both figures highlight the misconceptions of \citeauthor{Barnum2011}'s (\citeyear{Barnum2011}) proposal. It benchmarks the data points relative to the green line with the slope of $-1$ in all scenarios. This observation is entirely in line with our expectations, as we have established this conclusion above that \citeauthor{Barnum2011}'s (\citeyear{Barnum2011}) proposal would lead to exactly the same result if perfect substitutability was assumed. Clearly, it does not capture the correct shape of the input isoquant, even if we have perfectly accurate data with zero noise and inefficiency (as shown in panels \ref{fig30:a} and \ref{fig1000:a}). In contrast, both CCR DEA and FP-constrained DEA achieve a perfect fit to the true input isoquant in the case of no inefficiency.

Combining Figures \ref{fig:sample30} and \ref{fig:sample1000}, when inefficiencies are present, one can correctly argue that CCR DEA produces a wrong shape of the input isoquant with a small sample size (e.g., 30), yet the CCR DEA frontier converges to the true frontier when the sample size is sufficiently large (e.g., 1000). In contrast, FP-constrained DEA can mimic the correct shape of the true frontier even in a small sample and further, it performs consistently better than CCR DEA.

\section{Conclusions}\label{sec:concl}
\textcite{Barnum2011} and \textcite{Barnum2017} argue that DEA is inappropriate for estimating fixed proportion technologies because DEA implicitly assumes inputs are substitutable and outputs are transformable. In fact, nothing in the classical axioms of DEA limits substitutability or transformability. We agree with the conceptual point that if the true frontier of the underlying technology is L-shaped, the DEA frontier might have a wrong shape. That being said, this issue is merely finite sample bias. Going back to the classic works in economics by \textcite{Koopmans1951} and \textcite{Afriat1972} and the DEA literature later developed by \textcite{Banker1993} and \textcite{simar_aspects_1996}, we know that DEA is statistically consistent under axioms A1--A3. Since the fixed proportion technology does satisfy all the axioms, DEA is consistent in this case.

In this paper, we have proposed the correct modeling of fixed proportion technologies
within the axiomatic framework. A Monte Carlo study confirms the finite sample bias in the DEA modeling of fixed proportion technologies and the superiority of our proposed FP-constrained DEA over the original CCR DEA. If one does not have prior information about the substitution or transformation possibilities but the assumptions of DEA hold, then it is better to use standard DEA because it is consistent for the entire range between perfect substitution (transformation) as one extreme case and non-substitution (non-transformation) as the other extreme. The finite sample bias can be alleviated by simply increasing the sample size. However, if one needs to make an additional assumption on the substitution or transformation possibilities, FP-constrained DEA can more accurately model fixed proportion technologies not only in large samples but notably in small ones. 

We believe that this paper opens interesting avenues for future research. The proposed FP-constrained DEA can readily be extended to various DEA models. It is possible to generalize the fixed proportion axioms to one or more arbitrary subsets of non-substitutable inputs and/or non-transformable outputs. Such subsets might also include undesirable outputs and thus, it would be relevant to consider alternative axioms such as weak disposability. Extension to convex regression and related techniques \parencite{Kuosmanen2008,Kuosmanen2012} is also left as an interesting challenge for future research.

\section*{Acknowledgments}\label{sec:ack}

We are grateful for computational support from the University of York High Performance Computing service, Viking and the Research Computing team. The earlier versions of this paper were presented at the 2018 INFORMS Annual Meeting, the 17th European Workshop on Efficiency and Productivity Analysis, the Asia Pacific Productivity Conference 2022, the DEA45: International Conference on Data Envelopment Analysis, and the DEA Workshop on Performance Analytics, AI, and Sustainability. We thank Victor Podinovski, Sebastián Lozano, Léopold Simar, John Ruggiero, Ali Emrouznejad, Dimitris Giraleas, Mehdi Toloo, and other participants for their helpful comments.


\printbibliography

@article{kuosmanen_dea_2001,
	series = {Data {Envelopment} {Analysis}},
	title = {{DEA} with efficiency classification preserving conditional convexity},
	volume = {132},
	issn = {0377-2217},
	url = {https://www.sciencedirect.com/science/article/pii/S0377221700001557},
	doi = {10.1016/S0377-2217(00)00155-7},
	number = {2},
	urldate = {2024-12-20},
	journal = {European Journal of Operational Research},
	author = {Kuosmanen, Timo},
	month = jul,
	year = {2001},
	pages = {326--342},
}

@article{barnum_bias_2021,
	title = {Bias in {Transport} {Efficiency} {Estimates} {Caused} by {Misspecified} {DEA} {Models}},
	volume = {55},
	number = {4},
	journal = {Journal of Transport Economics and Policy (JTEP)},
	author = {Barnum, Darold and Coupet, Jason and Gleason, John and McWilliams, Abagail and Parhankangas, Annaleena},
	month = oct,
	year = {2021},
	pages = {356--373},
}

@book{leontief_structure_1941,
	address = {Cambridge Mass.},
	title = {The {Structure} of the {American} {Economy}, 1919–1929},
	publisher = {Harvard University Press},
	author = {Leontief, Wassily W.},
	year = {1941},
}

@article{diewert_application_1971,
	title = {An {Application} of the {Shephard} {Duality} {Theorem}: {A} {Generalized} {Leontief} {Production} {Function}},
	volume = {79},
	issn = {0022-3808},
	shorttitle = {An {Application} of the {Shephard} {Duality} {Theorem}},
	url = {https://www.jstor.org/stable/1830768},
	number = {3},
	urldate = {2024-11-06},
	journal = {Journal of Political Economy},
	author = {Diewert, W. E.},
	year = {1971},
	note = {Publisher: The University of Chicago Press},
	pages = {481--507},
}

@article{fare_structure_1983,
	title = {The {Structure} of {Technical} {Efficiency}},
	volume = {85},
	doi = {10.2307/3439477},
	journal = {Scandinavian Journal of Economics},
	author = {Färe, Rolf and Grosskopf, Shawna and Lovell, Ca},
	month = feb,
	year = {1983},
	pages = {181--90},
}

@article{mehdiloo_strong_2021,
	title = {Strong, weak and {Farrell} efficient frontiers of technologies satisfying different production assumptions},
	volume = {294},
	issn = {03772217},
	url = {https://linkinghub.elsevier.com/retrieve/pii/S0377221721000394},
	doi = {10.1016/j.ejor.2021.01.022},
	language = {en},
	number = {1},
	urldate = {2024-11-03},
	journal = {European Journal of Operational Research},
	author = {Mehdiloo, Mahmood and Podinovski, Victor V.},
	month = oct,
	year = {2021},
	pages = {295--311},
}

@article{Farrell1957,
	title = {The measurement of productive efficiency},
	volume = {120},
	number = {3},
	journal = {Journal of the Royal Statistical Society. Series A (General)},
	author = {Farrell, M. J.},
	year = {1957},
	pages = {253--290},
}

@article{tulkens_fdh_1993,
	title = {On {FDH} {Efficiency} {Analysis}: {Some} {Methodological} {Issues} and {Applications} to {Retail} {Banking}, {Courts}, and {Urban} {Transit}},
	volume = {4},
	journal = {Journal of Productivity Analysis},
	author = {Tulkens, Henry},
	year = {1993},
	pages = {183--210},
}

@article{alvarez_data_2020,
	title = {A {Data} {Envelopment} {Analysis} {Toolbox} for {MATLAB}},
	volume = {95},
	issn = {1548-7660},
	url = {http://www.jstatsoft.org/v95/i03/},
	doi = {10.18637/jss.v095.i03},
	language = {en},
	number = {3},
	urldate = {2023-08-11},
	journal = {Journal of Statistical Software},
	author = {Álvarez, Inmaculada C. and Barbero, Javier and Zofío, José L.},
	year = {2020},
}

@article{Barnum2011,
	title = {Measuring efficiency under fixed proportion technologies},
	volume = {35},
	issn = {0895-562X},
	url = {http://link.springer.com/10.1007/s11123-010-0194-y},
	doi = {10.1007/s11123-010-0194-y},
	number = {3},
	urldate = {2018-06-09},
	journal = {Journal of Productivity Analysis},
	author = {Barnum, Darold and Gleason, John},
	month = jun,
	year = {2011},
	note = {Publisher: Springer US},
	pages = {243--262},
}

@article{guner_multi-period_2021,
	title = {Multi-period efficiency analysis of major {European} and {Asian} airports under fixed proportion technologies},
	volume = {107},
	issn = {0967-070X},
	url = {https://www.sciencedirect.com/science/article/pii/S0967070X21001104},
	doi = {10.1016/j.tranpol.2021.04.015},
	language = {en},
	urldate = {2023-08-11},
	journal = {Transport Policy},
	author = {Güner, Samet and Cebeci, Halil İbrahim},
	month = jun,
	year = {2021},
	pages = {24--42},
}

@book{Fare1995,
	address = {Boston},
	title = {Multi-{Output} {Production} and {Duality}: {Theory} and {Applications}},
	isbn = {978-94-010-4284-0},
	url = {https://www.google.com/books?hl=en&lr=lang_en&id=WbXvCAAAQBAJ&oi=fnd&pg=PP8&dq=Multi-Output+Production+and+Duality:+Theory+and+Applications&ots=b-QoLK7vSy&sig=wyyrBLkeM3pDfW8WPPEUHmcbkCA},
	urldate = {2018-05-19},
	publisher = {Kluwer Academic Publishers},
	author = {Färe, Rolf and Primont, Daniel},
	year = {1995},
	pmid = {25246403},
	doi = {10.1007/978-94-011-0651-1},
	note = {arXiv: 1011.1669v3
ISSN: 1098-6596},
}

@article{Boyabatl2015,
	title = {Supply {Management} in {Multiproduct} {Firms} with {Fixed} {Proportions} {Technology}},
	volume = {61},
	issn = {0025-1909},
	url = {http://pubsonline.informs.org/doi/10.1287/mnsc.2014.2055},
	doi = {10.1287/mnsc.2014.2055},
	number = {12},
	journal = {Management Science},
	author = {Boyabatlı, Onur},
	year = {2015},
	note = {ISBN: 0025-1909},
	pages = {3013--3031},
}

@article{simar_aspects_1996,
	title = {Aspects of statistical analysis in {DEA}-type frontier models},
	volume = {7},
	issn = {0895562X},
	doi = {10.1007/BF00157040},
	number = {2-3},
	journal = {Journal of Productivity Analysis},
	author = {Simar, Léopold},
	year = {1996},
	pages = {177--185},
}

@article{Kuosmanen2008,
	title = {Representation theorem for convex nonparametric least squares},
	volume = {11},
	issn = {13684221},
	url = {http://doi.wiley.com/10.1111/j.1368-423X.2008.00239.x},
	doi = {10.1111/j.1368-423X.2008.00239.x},
	number = {2},
	urldate = {2018-05-27},
	journal = {Econometrics Journal},
	author = {Kuosmanen, Timo},
	month = jul,
	year = {2008},
	note = {Publisher: Wiley/Blackwell (10.1111)
ISBN: 1368-423X},
	pages = {308--325},
}

@article{Kuosmanen2012,
	title = {Stochastic non-smooth envelopment of data: {Semi}-parametric frontier estimation subject to shape constraints},
	volume = {38},
	issn = {0895562X},
	url = {http://link.springer.com/10.1007/s11123-010-0201-3},
	doi = {10.1007/s11123-010-0201-3},
	number = {1},
	urldate = {2018-05-21},
	journal = {Journal of Productivity Analysis},
	author = {Kuosmanen, Timo and Kortelainen, Mika},
	month = aug,
	year = {2012},
	note = {Publisher: Springer US
ISBN: 0895-562X{\textbackslash}r1573-0441},
	pages = {11--28},
}

@article{Banker1993,
	title = {Maximum {Likelihood}, {Consistency} and {Data} {Envelopment} {Analysis}: {A} {Statistical} {Foundation}},
	volume = {39},
	issn = {0025-1909},
	url = {http://pubsonline.informs.org/doi/abs/10.1287/mnsc.39.10.1265},
	doi = {10.1287/mnsc.39.10.1265},
	number = {10},
	urldate = {2018-07-11},
	journal = {Management Science},
	author = {Banker, Rajiv D.},
	month = oct,
	year = {1993},
	pages = {1265--1273},
}

@article{Charnes1978,
	title = {Measuring the efficiency of decision making units},
	volume = {2},
	issn = {03772217},
	url = {https://www.sciencedirect.com/science/article/pii/0377221778901388?via%3Dihub},
	doi = {10.1016/0377-2217(78)90138-8},
	number = {6},
	urldate = {2018-05-27},
	journal = {European Journal of Operational Research},
	author = {Charnes, A. and Cooper, W. W. and Rhodes, E.},
	month = nov,
	year = {1978},
	pmid = {7347358},
	note = {Publisher: North-Holland
ISBN: 03772217},
	pages = {429--444},
}

@article{Afriat1972,
	title = {Efficiency {Estimation} of {Production} {Functions}},
	volume = {13},
	issn = {00206598},
	doi = {10.2307/2525845},
	number = {3},
	urldate = {2020-04-06},
	journal = {International Economic Review},
	author = {Afriat, S. N.},
	month = oct,
	year = {1972},
	note = {Publisher: JSTOR},
	pages = {568--598},
}

@article{Allen1997,
	title = {Weights restrictions and value judgements in {Data} {Envelopment} {Analysis}: {Evolution}, development and future directions},
	volume = {73},
	issn = {02545330},
	doi = {10.1023/A:1018968909638},
	urldate = {2020-04-07},
	journal = {Annals of Operations Research},
	author = {Allen, R. and Athanassopoulos, A. and Dyson, R. G. and Thanassoulis, E.},
	year = {1997},
	note = {Publisher: Springer},
	pages = {13--34},
}

@article{banker1984,
	title = {Some {Models} for {Estimating} {Technical} and {Scale} {Inefficiencies} in {Data} {Envelopment} {Analysis}},
	volume = {30},
	issn = {0025-1909},
	url = {http://pubsonline.informs.org/doi/abs/10.1287/mnsc.30.9.1078},
	doi = {10.1287/mnsc.30.9.1078},
	number = {9},
	urldate = {2019-06-23},
	journal = {Management Science},
	author = {Banker, R. D. and Charnes, A. and Cooper, W. W.},
	month = sep,
	year = {1984},
	note = {Publisher: INFORMS},
	pages = {1078--1092},
}

@book{Shephard1970,
	address = {Princeton, New Jersey},
	title = {Theory of {Cost} and {Production} {Functions}},
	publisher = {Princeton University Press},
	author = {Shephard, Ronald W},
	year = {1970},
}

@article{Barnum2017,
	title = {Impact of input substitution and output transformation on data envelopment analysis decisions},
	volume = {49},
	issn = {0003-6846},
	url = {https://www.tandfonline.com/doi/full/10.1080/00036846.2016.1221042},
	doi = {10.1080/00036846.2016.1221042},
	number = {15},
	urldate = {2018-06-09},
	journal = {Applied Economics},
	author = {Barnum, Darold and Coupet, Jason and Gleason, John and McWilliams, Abagail and Parhankangas, Annaleena},
	month = mar,
	year = {2017},
	note = {Publisher: Routledge},
	pages = {1543--1556},
}

@incollection{Koopmans1951,
	address = {New York},
	title = {An analysis of production as an efﬁcient combination of activities},
	booktitle = {Activity {Analysis} of {Production} and {Allocation}},
	publisher = {Wiley},
	author = {Koopmans, Tjalling Charles},
	editor = {Koopmans, Tjalling Charles},
	year = {1951},
}
\baselineskip 12pt






\end{document}